\documentclass[fleqn,twoside,amsfonts]{article}

\makeatletter

\newcommand{\bls}[1]{\renewcommand{\baselinestretch}{#1}}

\def\noi{\noindent}

\renewcommand{\section}{\@startsection{section}{1}{0pt}%
        {-3.5ex plus -1ex minus -.2ex}{2.3ex plus .2ex}%
        {\large\bf\protect\raggedright}}

\renewcommand{\subsection}{\@startsection{subsection}{2}{0pt}%
        {-3ex plus -1ex minus -.2ex}{1.4ex plus .2ex}%
        {\normalsize\bf\protect\raggedright}}

\renewcommand{\thesubsubsection}%
        {\arabic{section}.\arabic{subsection}.\arabic{subsubsection}.}

\renewcommand{\@oddhead}{\raisebox{0pt}[\headheight][0pt]{%
   \vbox{\hbox to\textwidth{\rightmark \hfil \rm \thepage \strut}\hrule}}}
\renewcommand{\@evenhead}{\raisebox{0pt}[\headheight][0pt]{%
   \vbox{\hbox to\textwidth{\thepage \hfil \leftmark \strut}\hrule}}}
\newcommand{\heads}[2]{\markboth{\protect\small\it #1}{\protect\small\it
#2}}
\newcommand{\Acknow}[1]{\subsection*{Acknowledgement} #1}

\newcommand{\Title}[1]{\noi {\Large #1} \\}
\newcommand{\Author}[2]{\noi{\large\bf #1}\\[2ex]\noindent{\it #2}\\}

\newcommand{\Abstract}[1]{\vskip 2mm \begin{center}
        \parbox{16.4cm}{\small\noi #1} \end{center}\medskip}

\newcommand{\foom}[1]{\protect\footnotemark[#1]}

\newcommand{\email}[2]{\footnotetext[#1]{e-mail: #2}}

\def\nqq{\hspace*{-2em}}

\def\lal{&&\nqq {}}

\def\beq{\begin{equation}}
\def\eeq{\end{equation}}
\def\bear{\begin{eqnarray}}
\def\bearr{\begin{eqnarray} \lal}
\def\ear{\end{eqnarray}}
\def\earn{\nonumber \end{eqnarray}}
\def\nn{\nonumber\\ {}}

\makeatother

\def\beq#1{\begin{equation}\label{#1}}
\def\eeq{\end{equation}}
\def\ber#1{\begin{eqnarray}\label{#1} \nqq}
\def\eer{\end{eqnarray}}

\catcode`\@=11 \@addtoreset{equation}{section}\catcode`\@=12

\usepackage{amssymb}
\usepackage{latexsym}
\usepackage[dvips]{graphicx}
\newcommand{\be}{\begin{equation}}
\newcommand{\ee}{\end{equation}}
\newcommand{\bea}{\begin{eqnarray}}
\def\nn{\nonumber}
\newcommand{\eea}{\end{eqnarray}}

\heads{ T. Christodoulakis, E. Korfiatis, E.C. Vagenas}
     {Simultaneous Hamiltonian Treatment
of Class A Spacetimes and... }

\bls{0.98}
\begin{document}

\thispagestyle{empty}

\Title{Simultaneous Hamiltonian Treatment \\
of Class A Spacetimes\\ and Reduction of Degrees of Freedom at the
Quantum Level}

\Author{T. Christodoulakis\foom 1 ,
 E. Korfiatis, E.C. Vagenas\foom 2} {University of Athens, Physics
Department, Nuclear and Particle Physics Section,\\
Panepistimioupolis, Ilisia 157 71, Athens, Greece}
\\
\\{\it {\bf Dedication}\\
One of us (T.C.) wishes to dedicate this work to the memory of his
late father Ioannis Christodoulakis whose sudden death during the
preparation of an early version of this work in $1994$ greatly
sorrowed him.}

\Abstract {We consider the quantization of a general
spatially homogeneous space-time belonging to an arbitrary but
fixed Class A Bianchi type. Exploiting the information furnished
by the quantum version of the momentum constraints, we use as
variables the two simplest scalar contractions of
$C^{\alpha}_{\beta \gamma}$ and $\gamma_{\alpha \beta}$ as well as
the determinant of $\gamma_{\alpha \beta}$ ; We thus arrive at an
equation for the wave function in terms of these quantities. This
fact enables us to treat in a uniform manner all Class A
cosmologies. For these spacetimes the Langrangian used correctly
reproduces Einstein's equations. We also discuss the imposition of
 simplifying ans\"{a}tzen at the quantum level in a way that respects
the scalings of the quadratic Hamiltonian constraint.}

\email 1 {tchris@cc.uoa.gr} \email 2{hvagenas@cc.uoa.gr}

 \section{Introduction}

 Since the pioneering work of
DeWitt \cite{dewitt} a lot of minisuperspace models have appeared
in the literature \cite{haliwell}. In all these models, a
particular Bianchi type symmetry group is selected and a, more or
less, simple form for the time dependent matrix $\gamma_{\alpha
\beta}(t)$ (appearing in the spatial part of the metric) is
adopted; it is not thus easy to determine which of the properties
of these quantum models can be expected to hold as more and more
degrees of freedom (eventually leading to the full theory) are
taken into account. Yet, the investigation of precisely such kind
of properties is one of the main reasons for considering Quantum
Cosmology. In this work we try to avoid this fragmentation by
considering the wavefunction to be a function of the minimum
possible number of scalar under the action GL(3,R) combinations of
$C^{\alpha}_{\beta \gamma}$ and $\gamma_{\alpha \beta}$, and
$\gamma_{\alpha \beta}$ to be (in principle) unrestricted. The
paper is organised as follows : In section II we set-up the
Wheeler-DeWitt
 equation for a general spatially homogeneous metric.
Thus, exploiting the relations between the curvature
invariants and the ``scalars'' constructed out of
 $C^{\alpha}_{\beta \gamma}$ and $\gamma_{\alpha \beta}$
  we present an equation for $\Psi$ in terms of the two simplest
 independent scalar contractions and $\gamma=det\gamma_
{\alpha \beta}$; As an indicative example we solve
the resulting equation for the type II case. In section III
 we describe a reduction scheme for implementing various
simplifing ans\"{a}tzen at the quantum level. Employing two
 simple
ans\"{a}tzen we again give, using this reduction scheme, two
indicative examples of reduced minisuperspace models.
Finally the results obtained are discussed in section IV.
\section{Curvature Invariants and
The Reduced Wheeler-DeWitt Equation}

Our starting point is the action for pure gravity
\be
S_{gr}=-\int d^{3}xdt\sqrt{-{}^{(4)}g}\:{}^
{(4)}R
\label{2.1}
\ee
A spatially homogeneous space-time is characterised
 by the line element
\be ds^{2}=-N^{2}(t)dt^{2}+g_{ij}(x,t)dx^{i}dx^{j}+2N_{i} dx^{i}dt
\label{2.2} \ee where \be g=\gamma_{\alpha
\beta}(t)\sigma^{\alpha}(x) \sigma^{\beta}(x)\Leftrightarrow
g_{ij}= \gamma_{\alpha \beta}(t)\sigma^{\alpha}_{i}(x)
\sigma^{\beta}_{j}(x) \label{2.3a} \ee \be
N_{i}=N_{\alpha}(t)\sigma^{\alpha}_{i}(x) \label{2.3aa} \ee Lower
case Latin indices are world (tensor) indices and range from
1$\ldots$3 while lower case Greek indices number the different
one-forms and take values in the same range. The exterior
derivative of any basis one-form (being a two-form) is a linear
combination of the wedge product of any two of them, i.e. \be
d\sigma^{\alpha}=C^{\alpha}_{\beta \gamma}\;\sigma^{\beta}
\wedge\sigma^{\gamma}\Leftrightarrow
\sigma^{\alpha}_{i,j}-\sigma^{\alpha}_{j,i}= 2C^{\alpha}_{\beta
\gamma}\sigma^{\gamma}_{i} \sigma^{\beta}_{j} \label{2.3b} \ee The
coefficients $C^{\alpha}_{\beta \gamma}$ are,
 in general, functions of the point $x$. When the
space is homogeneous and admits a $3$-dimensional isometry group,
there exist $3$ one-forms such that the $C$'s become independent
of $x$ and are then called structure constants of the
corresponding isometry group. After substitution of metric
(\ref{2.2}) into action (\ref{2.1}), the passage to the phase
space leads to the following Hamiltonian action \cite{maccallum} :
\be S=\int dt\left[N_{7}\left(\frac{1}{2}G_{\alpha \beta \gamma
\delta}\pi^{\alpha \beta}\pi^{\gamma \delta} +\sqrt{\gamma}R
\right)-1N^{\rho}\left( C^{\alpha}_{\beta \rho}\gamma_{\alpha \mu}
\pi^{\beta \mu}+C^{\gamma}_{\mu \gamma} \gamma_{\rho
\epsilon}\pi^{\epsilon \mu}\right)\right] \label{2.4} \ee where
\be G_{\alpha \beta \gamma \delta}=\gamma^{-1/2}
\left(\gamma_{\alpha \gamma}\gamma_{\beta \delta} +\gamma_{\alpha
\delta}\gamma_{\beta \gamma}- \gamma_{\alpha \beta}\gamma_{\gamma
\delta} \right), \hspace{1.5cm}\gamma=det\left[\gamma_{\alpha
\beta} \right] \label{2.5a} \ee with inverse \be G^{\alpha \beta
\gamma \delta}=\frac{1}{4}\gamma^{1/0} \left(\gamma^{\alpha
\gamma}\gamma^{\beta \delta} +\gamma^{\alpha \delta}\gamma^{\beta
\gamma}- 2\gamma^{\alpha \beta}\gamma^{\gamma \delta} \right)
\label{2.7aa} \ee \be N^{2}_{0}=-N^{2}+\gamma^{\alpha
\beta}N_{\alpha} N_{\beta}, \hspace{1.5cm}N^{\rho}=\gamma^{\rho
\delta}N_{\delta} \label{2.5aaa} \ee and \be R=C^{\beta}_{\lambda
\mu}C^{\alpha}_{\theta \tau} \gamma_{\alpha \beta}\gamma^{\theta
\lambda} \gamma^{\tau \mu}+2\gamma^{\beta \nu} C^{\alpha}_{\beta
\delta}C^{\delta}_{\nu \alpha}+ 4\gamma^{\nu \lambda}C^{\mu}_{\mu
\nu}C^{\beta}_ {\beta \lambda} \label{2.5b} \ee is the Ricci
scalar of the slice t=constant. Action (\ref{2.4}) describes a
system of ten degrees of freedom (six $\gamma_{\alpha \beta}$ plus
the Lagrange multipliers $N_{0}$, $N_{\alpha}$) among which there
are constraints. Treating this system according to Dirac's
\cite{dirac} theory one arrives at the primary constraints
$\left(\Pi^{0},\Pi^{\alpha} \right)\approx(0,0,0,0)$ which in turn
lead to the secondary : \be H_{0}:=\frac{1}{2}G_{\alpha \beta
\gamma \delta} \pi^{\alpha \beta}\pi^{\gamma
\delta}+\sqrt{\gamma}R\approx 0
\hspace{1cm}H_{\alpha}:=C^{\nu}_{\beta \alpha}\gamma _{\nu
\mu}\pi^{\beta \mu}+C^{\gamma}_{\mu \gamma} \gamma_{\alpha
\tau}\pi^{\mu \tau}\approx 0. \label{2.6a} \ee If we restrict
attention to the Class A subgroup then $C^{\gamma}_{\mu \gamma}=0$
and constraints (\ref{2.6a}) are first class, satisfying : \be
\left\{H_{0},H_{\alpha}\right\}=0,\hspace{1cm}
\left\{H_{\alpha},H_{\beta}\right\}=-\frac{1}{2}
C^{\gamma}_{\alpha \beta}H_{\gamma}. \label{2.6b} \ee In deriving
(\ref{2.6b}) use has been made of the basic P.B.'s\quad
$\left\{\gamma_{\alpha \beta} ,\pi^{\mu \nu}\right\}=\delta^{\mu
\nu}_{\alpha \beta}$. For the Class B case further restrictions on
the phase- space variables are introduced by the requirement that
constraints (\ref{2.6a}) should be preserved in time, indicating
that action (\ref{2.4}) does not correctly reproduces Einstein's
Equations \cite{maccallum}. For this reason it is understood that
in the rest of the paper we are mostly concerned with Class A
spacetimes.
\par\noindent
In order to quantize the classical action (\ref{2.4})
we follow Dirac's quantization prescription \cite{dirac} : Firstly
we have to realize constraints (\ref{2.6a})
 as (possibly Hermitian) operators acting on some
Hilbert space spanned by their common null eigenvectors;
 secondly, we must check consistency, i.e. we need to
verify that these operators form a closed quantum algebra,
preferably isomorphic to (\ref{2.6b}); thirdly, we have
to find the $\Psi$'s by solving the linear quantum
 constraints as well as the Wheeler-DeWitt equation. Adopting
 the Schr\"{o}dinger representation, i.e. :
\be \hat{\gamma}_{\alpha \beta}=\gamma_{\alpha
\beta},\hspace{0.5cm} \hat{\pi}^{\alpha
\beta}=-i\frac{\partial}{\partial \gamma_{ \alpha
\beta}},\hspace{0.5cm}
\left(\hat{N}_{0},\hat{N}_{\alpha}\right)=\left(N_{0},N_{\alpha}\right)
\hspace{0.5cm}\left(\hat{\Pi}^{0},\hat{\Pi}^{\alpha}\right)=
\left(-i\frac{\partial}{\partial N_{0}},-i\frac{\partial}
{\partial N_{\alpha}}\right), \ee we readily see that the quantum
analogs of the primary constraints simply inform us that $\Psi$ is
not a function
 of $N_{0}$, $N_{\alpha}$.
In turning the linear constraints (\ref{2.6a}) into operators we
adopt the prescription of having all the $\pi^{\alpha \beta}$'s to
the far right so that the quantum constraints maintain the
classical symmetries. Thus we have : \be
\hat{H}_{\alpha}\Psi:=C^{\nu}_{\beta \alpha}\gamma _{\nu
\mu}\hat{\pi}^{\beta \mu}\Psi=0 \label{2.7b} \ee In turning the
quadratic constraint (\ref{2.6a}) into an operator annihilating
the wave function $\Psi$ we adopt the quantization prescription of
Kuchar and Hajicek \cite{kuchar}, which amounts in realizing the
``kinetic part'' of this constraint as the conformal Laplacian
based on the ``physical'' metric :
\begin{eqnarray*}
\Sigma^{A B}=\frac{\partial x^A}{\partial \gamma_{\alpha \beta}}
\frac{\partial x^B}{\partial \gamma_{\gamma \delta}} G_{\alpha
\beta \gamma \delta}
\end{eqnarray*}
\par \noindent
induced by $G_{\alpha \beta \gamma \delta}$ and the solutions
$x^{A}$ to the linear quantum constraints (\ref{2.7b}) : \be
\nabla_{\Sigma}^{Conf}=\nabla_{\Sigma}+\frac{D-2}{4\left(D-1\right)}\mathcal{R}_{\Sigma}
\ee D is the dimension of the reduced minisuperspace spanned by
the $x^{A}$'s.
 This quantization rule is mandatory if
one wants to respect the classical covariance of the action
(\ref{2.4}) :
\be
\gamma_{\alpha \beta}\rightarrow \tilde{\gamma}_{\alpha \beta}
(\gamma_{\mu \nu}),\hspace{1cm} N\rightarrow \tilde{N}=
f(\gamma_{\alpha \beta})N,
\ee
or, equivalently, the freedom to multiply constraint $H_{0}$
 by an arbitrary function.
\par
\noindent Thus, the Wheeler-DeWitt equation which $\Psi$ has
 to satisfy takes the form :
\be -\frac{1}{2}\left[ \Sigma^{A B}\frac{\partial^{2}}{\partial
x^{A}\partial x^{B}} -\Gamma^{\Lambda}_{A B}\Sigma^{K
B}\frac{\partial}{\partial
x^{\Lambda}}+\frac{D-2}{4(D-1)}\mathcal{R}_{\Sigma}+\sqrt{\gamma}R
\right]\Psi=0 \label{2.7a} \ee
\par \noindent
where $\Gamma^{\Lambda}_{A B}$, $\mathcal{R}_{\Sigma}$ are the
Christoffel symbol's and the Ricci scalar corresponding to $\Sigma
_{M N}$ (inverse to $\Sigma^{A B}$). It is tedious but
straightforward to verify (using
 $\left[\hat{\gamma}_{\alpha \beta},\hat{\pi}^{\gamma
\delta}\right]=-i\delta^{\gamma \delta}_{\alpha \beta}$) that the
quantum constraints satisfy an algebra completely analogous to
(\ref{2.6b}) : \be \left\{\hat{H}_{0},\hat{H}_{\alpha}\right\}=0,
\hspace{0.5cm} \left\{\hat{H}_{\alpha},\hat{H}_{\beta}\right\}=
-\frac{i}{2}C^{\gamma}_{\alpha \beta}\hat{H}_{\gamma} \label{2.7c}
\ee We next come to the measure of the Hilbert space to be
constructed by the solutions to (\ref{2.7a}, \ref{2.7b}) : It is
desirable to have a measure $\mu$ under which the
state-vector-defining operators (\ref{2.7a}, \ref{2.7b}) be
Hermitian; the most general Hermitian operator conditions
corresponding to the linear constraints (\ref{2.6a}) are
\cite{theo1} : \be \hat{X}_{\alpha}\Psi=\frac{1}{2\mu}\left(\mu
C^{\beta} _{\alpha \gamma}\gamma_{{}_{\beta
\delta}}\hat{\pi}^{\gamma \delta}+\hat{\pi}^{\gamma
\delta}C^{\beta}_{\alpha \gamma} \gamma_{{}_{\beta \delta}}
\mu\right)\Psi=0 \label{8.7d} \ee If we want $\hat{X}_{\alpha}$'s
to be identical to $\hat{H}_{\alpha}$'s in (\ref{2.7b}) we
immediately conclude (bearing in mind that $C^{\tau}_{\tau
\alpha}=0$) that $\mu$ must also satisfy (\ref{2.7b}), i.e. : \be
C^{\beta}_{\alpha \gamma}\gamma_{\beta \delta}\hat{\pi}^{ \gamma
\delta}\mu=0 \label{2.7e} \ee Finally, the Hermiticity of
(\ref{2.7a}) fixes the measure to be : \be \mu=\sqrt{det\Sigma_{A
B}} \label{2.7f} \ee which can be seen to satisfy (\ref{2.7e}).
\par
\noindent Let us now start our investigation of the space of state
vectors by considering the solutions to (\ref{2.7b}): Except for
the trivial Bianchi type I (see end of the section IV) and type II
(which is explicitly solved below), in all other Class
 A Bianchi types conditions (\ref{2.7b}) comprise a set
of $3$ independent PDE's in $6$ variables. As a result, we
expect the general solution to these equations, if it exists,
to be a function of $3$ combinations $\gamma_{\alpha \beta}$;
the existence of the solutions to (\ref{2.7b}) is guaranteed
 by Frobenius theorem : The quantum algebra (\ref{2.7c}) satisfied
 by the $\hat{H}_{\alpha}$'s (seen as vector fields on the
configuration space) is essentially the integrability
 conditions required by the theorem. As one can easily
verify, every combination of $C^{\alpha}_{\beta \gamma}$ and
$\gamma_{\alpha \beta}$ that is scalar under the action GL($3$,R)
(i.e. has all the frame indices contracted), as well as $\gamma$
is a solution to these equations. Thus the general solution to
(\ref{2.7b}) is a function of $\gamma$ and any scalar combination
of $C^{\alpha}_{\beta \gamma}$ and $\gamma_{\alpha \beta}$.
\par
\noindent
As it is well known \cite{weinberg}, the number of independent
curvature invariants of a three-dimensional space is
generically three. One can arbitrarily choose them to be :
\be
R_{1}=g^{ij}R_{ij}\hspace{1cm}
R_{2}=R^{i}_{j}R^{j}_{i}\hspace{1cm}
R_{3}=R^{i}_{j}R^{j}_{k}R^{k}_{i}
\ee
 For the case of (the spatial part of) metric (\ref{2.2}) :
\bea R^{ij}&=&\sigma^{i}_{\kappa}\sigma^{j}_{\lambda}
\left(C^{\kappa}_{\beta \delta}C^{\lambda}_{\mu \nu}
\gamma^{\delta \mu}\gamma^{\beta \nu}+2C^{\alpha} _{\beta
\gamma}C^{\gamma}_{\mu\alpha}\gamma^{\beta \kappa} \gamma^{\mu
\lambda}-2C^{\kappa}_{\beta \delta} C^{\nu}_{\mu
\nu}\gamma^{\delta \mu}C^{\gamma}_{\mu \alpha}
\gamma^{\beta \lambda}\right.\nn\\
&  &-\left. 2C^{\lambda}_{\beta \delta} C^{\nu}_{\mu
\nu}\gamma^{\delta \mu}\gamma^{\beta \kappa} +2C^{\alpha}_{\beta
\delta}C^{\tau}_{\mu \nu} \gamma_{\alpha \tau}\gamma^{\nu
\delta}\gamma^{\beta \lambda} \gamma^{\mu \kappa}\right)
\label{2.8} \eea see \cite{theo2}. From this relation it is
evident that the curvature invariants will be complicated
combinations of terms involving scalar contractions of
$C^{\alpha}_{\beta \gamma}$ and $\gamma_{\alpha \beta}$ (and thus
they will also satisfy (\ref{2.7b})). Searching for the most
convenient contractions to take as independent variables we
immediately spot the
 three simplest contractions appearing in expression (\ref{2.5b})
for the Ricci scalar, namely :
\begin{eqnarray*}
C^{\beta}_{\lambda \mu}
C^{\alpha}_{\theta \tau}\gamma_{\alpha \beta}\gamma^{\theta \lambda}
\gamma^{\tau \mu},\hspace{1cm}\gamma^{\beta \nu}C^{\alpha}_{\beta \delta}
C^{\delta}_{\nu \alpha},\hspace{1cm}\gamma^{\nu \lambda}C^{\mu}_{\mu \nu}
C^{\beta}_{\beta \lambda}.
\end{eqnarray*}
However, as one can see in the appendix, the third contraction is a multiple
of the second i.e. $\gamma^{\nu \lambda}C^{\mu}_{\mu \nu}
C^{\beta}_{\beta \lambda}=\lambda\gamma^{\beta \nu}C^{\alpha}
_{\beta \delta}C^{\delta}_{\nu \alpha}$. The presence of $\gamma$
in the potential term of (\ref{2.7a}), as well as the fact that $\gamma$
is a solution to (\ref{2.7b}), suggests that the three independent variables
can be taken to be :
\be
x^{1}:=C^{\beta}_{\lambda \mu}C^{\alpha}_{\theta \tau}
\gamma_{\alpha \beta}
\gamma^{\theta \lambda}\gamma^{\tau \mu}\hspace{1cm}
x^{2}:=\gamma^{\beta \nu}C^{\alpha}_{\beta \delta}
C^{\delta}_{\nu \alpha}\hspace{1cm}
x^{3}:=\gamma
\label{2.9}
\ee
instead of $R_{1},\:R_{2},\:R_{3}$ (see appendix for the exact expressions
of $R_{1},\:R_{2},\:R_{3}$ in terms of $x^{1},\:x^{2},\:x^{3}$). Therefore
the general solution to (\ref{2.7b}) is an arbitrary function of
$x^{1},\:x^{2},\:x^{3}$ say $\Psi=\Psi(x^{1},x^{2},x^{3})$. Note that
$x^{1},\:x^{2},\:x^{3}$ are the ``physical coordinates''
defined by Kuchar and Hajicek \cite{kuchar}.
We now turn our attention to (\ref{2.7a}).
In trying to calculate $\Sigma^{A B}$  and thus implement
the conclusion that $\Psi=\Psi(x^{1},x^{2},x^{3})$ one is coming across
terms quartic in $C$'s which are not manifest combinations of
sums of products of $x^{1},\:x^{2},\:x^{3}$. One such typical term
could, for example, be :
\begin{eqnarray*}
\frac{\partial^{2}\Psi}{\partial x^{2}\partial x^{1}}
\frac{\partial x^{2}}{\partial \gamma_{\alpha \beta}}
\frac{\partial x^{1}}{\partial \gamma_{\gamma \delta}}
\gamma_{\alpha \gamma}\gamma_{\beta \delta}=
\frac{\partial^{2}\Psi}{\partial x^{2}\partial x^{1}}
(2C^{\kappa}_{\gamma \nu}C^{\nu}_{\alpha \kappa}C^{\epsilon}
_{\mu \delta}C^{\xi}_{\tau \beta}\gamma_{\epsilon \xi}
\gamma^{\gamma \mu}\gamma^{\alpha \tau}\gamma^{\delta \beta}-
C^{\kappa}_{\gamma \nu}C^{\nu}_{\alpha \kappa}
C^{\gamma}_{\mu \delta}C^{\alpha}_{\tau \beta}
\gamma^{\mu \tau}\gamma^{\delta \beta})
\end{eqnarray*}
In deriving this relation use has been made of the
identity :
\begin{eqnarray*}
\frac{\partial \gamma^{\alpha \beta}}
{\partial \gamma_{\gamma \delta}}=-\frac{1}{2}
\left(\gamma^{\alpha \gamma}\gamma^{\beta \delta}
+\gamma^{\alpha \delta}\gamma^{\beta \gamma}\right)
\end{eqnarray*}
easily obtainable from the basic relations :
\begin{eqnarray*}
\frac
{\partial \gamma_{\alpha \beta}}{\partial
 \gamma_{\gamma \delta}}=\delta^{\gamma
\delta}_{\alpha \beta},\quad\gamma^{\alpha \beta}
\gamma_{\beta \delta}=\delta^{\alpha}_{\delta}.
\end{eqnarray*}
In order to actually show that all terms of the sort are,
indeed, algebraic combinations of $x^{1},\:x^{2},\:x^{3}$
 we have to count the number of independent ``scalar''
contractions that can be constructed out of
$C^{\alpha}_{\beta \gamma}$ and $\gamma_{\alpha \beta}$.
The procedure we adopt is closely similar to that used in
\cite{weinberg} to count the number of independent
curvature invariants in d-dimensions. As it is well known,
$C^{\alpha}_{\beta \gamma}$, being the structure
constants of a 3-dimensional Lie group, are antisymmetric
in their lower indices and satisfy the Jacobi identities :
\begin{eqnarray*}
T^{\delta}_{\alpha \beta \gamma}:=C^{\delta}_{\mu \gamma}
C^{\mu}_{\alpha \beta}+C^{\delta}_{\mu \alpha}C^{\mu}_{\beta
\gamma}+C^{\delta}_{\mu \beta}C^{\mu}_{\gamma \alpha}=0
\end{eqnarray*}
$T^{\delta}_{\alpha \beta \gamma}$ being fully antisymmetric
in its lower indices. The symmetric matrix $\gamma_{\alpha \beta}$
has 6 independent components. The antisymmetry of
$C^{\alpha}_{\beta \gamma}$ leaves it with 9 components restricted
by the 3 independent Jacobi Identities. Thus $C^{\alpha}_{\beta \gamma}$
has 6 independent components. If we are interested in scalar contractions
of  $C^{\alpha}_{\beta \gamma}$ and $\gamma_{\alpha \beta}$ then the
9 components $\Lambda^{\alpha}_{\beta}$ of an element of GL(3,R) can
be chosen arbitrarily. Hence, the 6 independent components
of $\gamma_{\alpha \beta}$ and the 6 independent
$C^{\alpha}_{\beta \gamma}$'s are subject to 9 restrictions
when put together to form scalar combinations. The number
of independent such scalar combinations is therefore
$6+6-9=3$.
Note that 3 is the maximum number of independent scalars; it is
achieved only in Bianchi types VIII, IX : In these types
$R_{1}\not=R_{2}\not=R_{3}\not=R_{1}\not=0$ (this is why sometimes
it is said that these types mimic more faithfully the generic three-space);
and there is a number density of weight $-1$, namely
$m=det[m^{\alpha \beta}]$ (see appendix). This m can be used to promote
 $x^{3}$ to a true scalar by taking the combination $\frac{x^{3}}{m^{2}}$ .
In all other Bianchi types $m=0$ and, as it can be seen either by
direct calculation or from the relations given in the appendix,
the number of independent $R_{1},R_{2},R_{3}$ is less than 3; namely
it is two for types VI$_{-1}$, VII$_{0}$ one for type II and 0 for
type I and are sufficient to describe the corresponding
three geometries  \cite{theo9}.
 Now $x^{3}$ cannot be promoted to a true scalar.
Nevertheless, since it explicitly appears in the interaction term
of (\ref{2.7a}) and because it is a solution to (\ref{2.7b}), we must
include it among the independent variables.
\par
\noindent
Since m is a number (and thus transparent to the derivatives with
respect to $\gamma_{\alpha \beta}$) we conclude that (\ref{2.9}) are
valid variables for all Class A types. In any case the above
considerations show that the system of (\ref{2.7a})-(\ref{2.7b}) is
self consistent, as one expects from the first class algebra (\ref{2.7c}).
The same conclusion follows from the considerations of Kuchar and Hajicek
\cite{kuchar} for the particular case of zero cocycle.
\par
\noindent
Thus we finally get from (\ref{2.7a}) the equation :
\bea
(5x^{1}x^{1}+16(\lambda-1)x^{2}x^{2}-8\lambda x^{1}x^{2}
+128W)
\frac{\partial^{2} \Psi}{\partial x^{1}\partial x^{1}}
+(x^{2}x^{2}-16W)\frac{\partial^{2}\Psi}{\partial x^{2}
\partial x^{2}}\nn\\
+(-3x^{3}x^{3})\frac{\partial^{2}\Psi}
{\partial x^{3}\partial x^{3}}
+(2x^{1}x^{2}+16W)\frac{\partial^{2}\Psi}
{\partial x^{1}\partial x^{2}}+(2x^{1}x^{3})
\frac{\partial^{2}\Psi}{\partial x^{1}
\partial x^{3}}
+(2x^{2}x^{3})\frac{\partial^{2}\Psi}
{\partial x^{2}\partial
x^{3}}\nn\\
A^{\Lambda}\frac{\partial \Psi}{\partial x^{\Lambda}}
-2x^{3}\left(x^{1}+2(1+2\lambda) x^{2}\right)\Psi=0\hspace{3cm}
\label{2.10} \eea
 \be \mbox{where}\hspace{5.5cm}W:=\epsilon
m\sqrt{\frac{x^{1}-2x^{2}} {2x^{3}}}\hspace{4.5cm} \ee see
appendix for the definition of m, $\epsilon$ and
$A^{\Lambda}=-\Gamma^{\Lambda}_{M N}\Sigma^{M N}$ is the
corresponding linear term of the Laplacian. For all Bianchi types
except VIII, IX this term is $A^{\Lambda}=(5x^{1}-4\lambda x^{2},
x^{2}, -3x^{3})$. Equation (\ref{2.10}) can be  compactly written
as : \be \left[\Sigma^{A B}\partial_{A}\partial_{B}
+A^{\Lambda}\partial_{\Lambda}-2x^{3}
\left(x^{1}+2(1+2\lambda)x^{2}\right) \right]\Psi=0 \label{2.6'}.
\ee Note that using the conformal covariance of our operator we
have multiplied by $(x^{3})^{1/2}$ and thus the ``physical''
metric in (\ref{2.6'}) is :
\begin{eqnarray*}
\Sigma^{A B}=L_{\mu \nu \kappa \lambda}\frac{\partial x^{A}}
{\partial \gamma_{\mu \nu}}\frac{\partial x^{B}}{\partial
\gamma_{\kappa \lambda}}\quad\mbox{where}\;L_{\mu \nu \kappa \lambda}
=\gamma_{\mu \kappa}\gamma_{\nu \lambda}+\gamma_{\mu \lambda}
\gamma_{\nu \kappa}-\gamma_{\mu \nu}\gamma_{\kappa \lambda}
\end{eqnarray*}
Of course, $\lambda$ is zero for all Class A cases. We only include
it to show that the above reduction from (\ref{2.7a}) to (\ref{2.10}) can
be made for Class B Bianchi types as well; this is also the
reason it appears in the examples at the end of the next section.
This equation constitutes our main result; it has, over (\ref{2.7a}-\ref{2.7f}),
the conceptual advantage of enabling us to treat in a uniform manner
 each and every particular Class A Bianchi type. It also has the technical
 advantage of a much more simpler interaction term, while the reduced
minisuperspace metric $\Sigma ^{A B}$  is no more complicated than
$G_{\alpha \beta \gamma \delta}$ appearing in (\ref{2.7a}-\ref{2.7f}) (except
of course for the square root term which, however, vanishes for all
 but VIII, IX Bianchi types). For any (Class A) Bianchi type $\Sigma^{A B}$
 in (\ref{2.6'}) is flat; one can easily see it for the lower
 Bianchi types (all except the VIII, IX) by transforming to
 coordinates :
  \be d^{1}:=\sqrt{\frac{x^{1}-2x^{2}}{x^{2}}}\hspace{1.5cm}
d^{2}:=(x^{2})^{3/8}(x^{3})^{-1/8}\hspace{1.5cm} d^{3}:=x^{2}x^{3}
\label{2.11a} \ee in which equation (\ref{2.10}) becomes :
\begin{eqnarray}
\left[(d^{1})^{2}-2(\lambda-2)\right]\frac{\partial^2
\Psi}{\partial (d^1)^2}+2
d^{2}d^{3}\frac{\partial^2 \Psi}{\partial d^{2}\partial d^{3}}+
d^{1}\frac{\partial \Psi}{\partial d^{1}}
\nn\\
-2d^{3}((d^{1})^{2}+4(\lambda+1))
\Psi=0\hspace{2cm}
\label{2.11b}
\end{eqnarray}
making explicit the flatness of $\Sigma^{A B}$.
\par
\noindent
We now give, as an indicative example, the general solution to
(\ref{2.10}) for the Bianchi II geometry. This Bianchi type is of some
importance since it can be considered as the highly anisotropic
 limit of type IX and it has recently been also treated in
 \cite{lidsey}. The only non-vanishing structure constants are
$C^{1}_{23}=-C^{1}_{32}=1$ and one can easily verify that the
scalar $x^{2}$ in (\ref{2.9}) is identically zero. Thus the
wave-function satisfying (\ref{2.10}) depends only on
$x^{1}:=\chi$ and $x^{3}:=\gamma$ reducing to the equation : \be
5\chi^2\frac{\partial^2 \Psi}{\partial \chi^2}-3\gamma^{2}
\frac{\partial^{2}\Psi}{\partial \gamma^{2}}+2\chi \gamma
\frac{\partial^{2}\Psi}{\partial \chi\partial \gamma}+
5\chi\frac{\partial\Psi}{\partial\chi}-3\gamma\frac{\partial\Psi}
{\partial \gamma}-2\gamma\chi\Psi=0. \label{2.12a} \ee Effecting
the transformation $v=\gamma \chi$, $u=\gamma^{3} \chi$ and
assuming $\Psi(u,v)=U(u)V(v)$ we get two ODE's : \bea
16u^{2}\frac{d^{2}U}{du^{2}}+16u\frac{dU}{du}-CU=0\\
4v^{2}\frac{d^{2}V}{dv^{2}}+4v\frac{dV}{dv}-
\left(C+2v\right)V=0
\label{2.12b}
\eea
where C is a separation constant.
\par
\noindent
The first of these is an Euler equation while the second can be easily
transformed
into a Bessel equation, further reduction to the one and only
 true degree of freedom for the type II \cite{theo9},
namely $\chi$, requires considerations
of the quantum version of some classical integrals of motion
\cite{theo11}.
At this point we deem it important to conclude
this section underlining the fact that no extra (by hand) assumption
has been used in arriving at (\ref{2.10}); the only
``arbitrariness'' involved is the factor ordering of the $\pi^{\alpha
\beta}$'s to the far right in (\ref{2.7b}). Thus the theory itself
naturally leads us to consider as equivalent any two hexads $\gamma^{(1)}
_{\alpha \beta}$, $\gamma^{(2)}_{\alpha \beta}$ iff they correspond to the
 same triplet $(x^{1},x^{2},x^{3})$ for a given Class A Bianchi type
under consideration. This is hardly a surprise since the linear
constraints $H_{\alpha}$ are the generators of the inner
automorphisms and therefore the two exads corresponding to the
same triplet will be inner automorphically related, i.e.
$\gamma^{(2)}_{\alpha \beta}=\Lambda^{\mu}
_{\alpha}\Lambda^{\nu}_{\beta}\gamma^{(1)}_{\mu \nu}$ with
$\Lambda\in\mbox{InAut[G]}$ \cite{jantzen}. The proper
investigation of this point and other related issues is too long
to be included here; it has been given in \cite{theo9}. Finally,
we wish to remark that
 the solutions to (\ref{2.10}) are intended to form the Hilbert space
 of the general spatially homogeneous Bianchi cosmology for each
and every Class A group type separately : no transitions between
different Class A Bianchi types can be considered; indeed the
structure constants appearing in (\ref{2.9}) are merely (mostly)
discrete parameters and not dynamical variables. If one wishes
 to consider such transitions one has to go deeper in Quantum Gravity,
beyond the Quantum Cosmology approximation.

\section{Reducing the Degrees of Freedom
at the Quantum Level}
\par
\noindent
Despite the simplification achieved in the previous
section, equation (\ref{2.10}) is still too difficult
to solve in its full generality. We would like thus to study
simplified models with less (than six) degrees of freedom.
 The way to do this, up to now, was to insert a particular choice for $\gamma_{
\alpha \beta}$ (say diag(a(t), b(t), c(t)) in the classical
 action (\ref{2.4}), and quantize the resulting (further) reduced
action. We, on the other hand, are already at the quantum level
having as our degrees of freedom the combinations of
$\gamma_{\alpha \beta}$ and $C^{\alpha}_{\beta \gamma}$ appearing
in (\ref{2.9}); it is thus natural to adopt as simplifying
Ans\"{a}tz a restriction on $x^{A}$ of the form $f(x^{A})=0$
which, given a particular Bianchi type, can be translated into a
choice for the form of $\gamma_{\alpha \beta}$. We are thus
presented with the task of defining the ``reduced equivalent''
 of equation (\ref{2.10}). Although there are standard
ways to project an equation on a submanifold of a given manifold
\cite{kobayashi} , we have to bear in mind that the metric
structure on the manifold spanned by $x^{A}$ (as well as that
corresponding to the manifold spanned by the $\gamma_{\alpha
\beta}$'s) is known only up to a rescaling : we are free to
multiply (\ref{2.10})
 by an arbitrary function of $x^{A}$. Due to this, the
``reduction rule'' we are going to adopt must respect the
conformal transformations of $\Sigma^{A B}$; in fact the need to
respect the scalings of $G_{\alpha\beta\gamma\delta}$ is the main
reason behind Kuchar's recipe for realizing the kinetic part of
$H_{0}$ as the conformal Laplacian based on $\Sigma_{A B}$. In
seeking to construct this procedure, let for a moment be general
and consider an equation of the form : \be \left[\Sigma^{A B}
\partial_{A}\partial_{B}+A^{\Gamma}
\partial_{\Gamma}+V(x^{A})\right]\Psi=0
\label{3.1}
\ee
on a D-dimensional space $\mathcal{M}$ spanned by $x^{A},\quad
A=1,2,\ldots, D$. We shall restrict our consideration to
the case where $\Sigma^{A B}$ is non-degenerate
(det$\Sigma^{A B}\not=0$) and the submanifold on which
we wish to define the reduced equivalent of (\ref{3.1}) is
 non-null, i.e. no one of its tangent vectors has zero length
 with respect to $\Sigma^{A B}$. We are to interpret $\Sigma^{A B}$
 as a contravariant metric on $\mathcal{M}$ but, due to the freedom
 to arbitrary scale (\ref{3.1}), we are only allowed to
use in our construction the conformal equivalence
 class a representative of which is $\Sigma^{A B}$.
Suppose now that we  are given the D-L restricting conditions
$f^{i}(x^{A})=0$,\quad $i=D, D-1,\ldots,L+1$ defining the
L-dimensional submanifold $\mathcal{N}$ on which we wish
to restrict (\ref{3.1}). Consider the (D-L) one forms
$\Phi^{i}:=\frac{\partial f^{i}(x^{\Gamma})}
{\partial x^{\Delta}}dx^{\Delta}$ spanning the subspace
orthogonal (with respect to $\Sigma ^{A B}$) to the
cotangent space of $\mathcal{N}$. The hypothesis that
$\mathcal{N}$ is non-null means that $\Sigma^{\Gamma \Delta}
\Phi^{i}_{\Gamma}\Phi^{i}_{\Delta}\not=0$ for all $i=D, D-1,
\ldots,D-L$ (no summation in i). Let us suppose that $f^{i}$ are
such that the vector fields $\xi_{i}:=\Sigma^{\Gamma \Delta}
\Phi^{i}_{\Delta}\partial_{\Gamma}$ are surface forming
 (i.e. $\left\{\xi_{i},\xi_{i}\right\}=C^{k}_{ij}\xi_{k}$
where $\left\{.,.\right\}$
stands for the Lie bracket). This implies that the system
of first order partial differential equations $\xi^{\Gamma}_{i}
\partial_{\Gamma}\Omega(x^{\Delta})=0$ admits L functionally
 independent solutions $q^{\alpha}(x^{\Gamma})$,
$\alpha=1,2,\ldots,L$, that is to say the general solution to this
system is of the form : $\Omega(x^{\Gamma})=\Omega(q^{\alpha})$.
Furthermore, no one of the $f^{i}$'s can be functionally dependent
on the $q^{\alpha}$'s since if $f^{i}=f^{i}(q^{\alpha})$ for some
i, then $\Sigma^{\Gamma
\Delta}\Phi^{i}_{\Gamma}\Phi_{\Delta}^{i}=0$
 contradicting the hypothesis that $\mathcal{N}$ is non null. Therefore
the D functions $x'^{\Gamma}=(q^{\alpha},f^{i})$ constitute a
regular coordinate system for $\mathcal{M}$, i.e. the Jacobian
matrix $\left[\partial x'^{\Gamma}/\partial x^{\Delta}\right]$
is non singular. Thus, via the inverse mapping theorem, we can always
(in principle) express $x^{\Gamma}$ as functions of $x'^{\Gamma}$.
\par \noindent
We are now ready to define the reduction scheme : We supplement
(\ref{3.1}) with the requirement that $\Phi$ does not change when
 we move along the integral curves of each $\xi_{i}$; thus we define
as reduction of (\ref{3.1}) the system of partial differential
equations :
\bea
L_{\xi_{i}}\Psi=0, \hspace{0.5cm}i=D,D-1,\ldots,L+1\label{3.2a}\\
\left[\Sigma^{A
B}\partial_{A}\partial_{B}+A^{\Gamma}\partial_{\Gamma}
+V(x^{A})\right]\Psi=0\label{3.2b}. \eea Let as see why this
system defines an equation on $\mathcal{N}$. As we explained above
(\ref{3.2a}) has a general solution
$\Psi=\Psi(q^{\alpha})$.Inserting this into (\ref{3.2b}) we get :
\be \Sigma^{A B}\frac{\partial q^{\alpha}}{\partial x^{A}}
\frac{\partial q^{\beta}}{\partial x^{B}}
\frac{\partial^{2}\Psi}{\partial q^{\alpha}\partial q^{\beta}}+
\left(\Sigma^{A B}\frac{\partial^{2}q^{\gamma}} {\partial
x^{A}\partial x^{B}}+A^{\Gamma}\frac{\partial q^{\gamma}}{\partial
x^{\Gamma}}\right)\frac{\partial \Psi} {\partial
q^{\gamma}}+V\Psi=0 \label{3.3} \ee where the coefficients are, in
principle, functions of all $x^{\Gamma}$. Using the fact that
$x^{\Gamma}=x^{\Gamma}(q^{\alpha},f^{i})$ and restricting on
$\mathcal{N}$ (i.e. setting $f^{i}=0$), we finally get an equation
on $\mathcal{N}$ : \be \left[\sigma^{\alpha
\beta}(q)\partial_{\alpha}\partial_{\beta}
+\alpha^{\gamma}(q)\partial_{\gamma}+V(q)\right]\Psi(q)=0
\label{3.4} \ee
 In the rest of the section we give examples of the
above reduction scheme exhibiting the solutions
for a large class of five dimensional minisuperspace models
 (except Bianchi types VIII, IX).
Starting from (\ref{2.11a}-\ref{2.11b}) let us reduce the number of independent
$\gamma_{\alpha \beta}$ to 5 by imposing the restriction
 (simplifying Ans\"{a}tz)
$d^{1}=const.\:(\Leftrightarrow x^{1}=\mu x^{2})$.
\par \noindent
Thus the defining equation for $\mathcal{N}$ is :
\be
f(d^{1},d^{2},d^{3}):=d^{1}-c=0
\label{3.5}
\ee
 Hence the vector $\xi$ is $\xi=\left((d^{1})^{2}-
2(\lambda-2)\frac{\partial}{\partial d^{1}}\right)$,
implying that (\ref{3.2a}) is satisfied by
$\Psi=\Psi(d^{2},d^{3})$. Inserting this $\Psi$ into
(\ref{2.11a}-\ref{2.11b}) and restricting to $\mathcal{N}$ we
arrive at the equivalent of (\ref{3.4}) :
\be
d^{2}\frac{\partial^{2}\Psi(d^{2}d^{3})}{\partial d^{2}
\partial d^{3}}-C\,\Psi(d^{2},d^{3})=0
\ee
with $C:=c^{2}+4(\lambda+1)$. Effecting the transformation :
\begin{eqnarray*}
d^2=e^{s^2},\qquad d^3=s^3
\end{eqnarray*}
we arrive at a Klein-Gordon equation (in lightcone coordinates) :
\be \frac{\partial^2 \Psi}{\partial s^2 \partial s^3}=C \Psi
\label{Klein} \ee with well-known space of solutions. A particular
solution can be reached by separation of variables in
(\ref{Klein}) and reads : \be
\Psi(d^{2},d^{3})=(d^{2})^{l'}exp\left[Cd^{3}/l'\right] \ee where
$l'$ is another separation constant.\par\noindent Another very
interesting simplifying Ans\"{a}tz is :
\begin{eqnarray*}
d^{3}=c'\,(:const.)\:\Leftrightarrow x^{2}x^{3}=const.\quad.
\end{eqnarray*}
Similarly to the previous case we arrive from (\ref{2.11b}) at the
Wheeler-DeWitt equation for $\Psi$ which has the form : \be
\left[(d^{1})^{2}-2(\lambda-2)\right]\frac{\partial^2
\Psi}{\partial (d^1)^2} +d^{1}\frac{\partial \Psi}{\partial d^{1}}
-2d^{3}((d^{1})^{2}+4(\lambda+1)) \Psi=0\hspace{2cm}
\label{2.11bb} \ee This equation is of the Matthieu Modified type
and is equivalent to equation (39) of \cite{theo12} (under the
identification $d^{1}=\sqrt{w-2}$). There , of course, the
reduction has been achieved by means of the quantum version of a
classical integral of motion and not as a simplifying Ans\"{a}tz.
\section{Discussion}

In this work we have investigated the quantization of the general
spatially homogeneous geometry for each Class A Bianchi type.
Counting the number of independent scalar contractions of
$\gamma_{\alpha \beta}$ and $C^{\alpha} _{\beta \gamma}$ we were
led to identify the minimum number of variables (i.e. degrees of
freedom) which $\Psi$ can be assumed to depend upon : $x^{1},
x^{2}, x^{3}$ defined in (\ref{2.9}). The very fact that this is
possible establishes , in our opinion, the conceptual (as well as
formal) connection of Quantum Cosmology with full Quantum Gravity
: Indeed, adopting the long standing belief that the
wave-functional of Quantum Gravity must be a functional of the
geometry, let us restrict ourselves within the class
 of smooth functionals \cite{theo3}, consider some general
solution to the full Wheeler-DeWitt equation of the form
$\Omega=\Omega(I)$ with $I=\int d^{3}x\sqrt{g}\:F\!\left(
R^{i}_{i},R^{i}_{j}R^{j}_{i},R^{i}_{j}R^{j}_{k}R^{k}_
{i}\right)$. If we now insert in $\Omega$ the particular
metric (\ref{2.2}-\ref{2.3b}) we see, because
of (\ref{2.8}) and the relations for $R_{1},\,R_{2},\,R_{3}$
given in the appendix, that indeed $\Omega$ becomes a
function of\, $x^{1},\,x^{2},\,x^{3}$ (forgetting, of
course, the (possibly infinite) constant $c=\int d^{3}x\,
\sigma(x)$ with $\sigma=det(\sigma^{\alpha}_{i})$ which
multiplies the $\sqrt{x^{3}}$).
\par
\noindent
An important new feature of (\ref{2.10}) is also the
fact that it offers the possibility to study the
quantum mechanics of spatially homogeneous geometries
in a uniform manner for all (Class A) Bianchi types.
Thus, the properties common to all these models can be
more easily revealed. Our inability to solve (\ref{2.10})
in its full generality leads us to consider simplified
models, but again the advantage over the usual
procedure is obvious : In our case the Ans\"{a}tz  is made
on the $x^{A}$'s enabling us to treat large subgroups
of simplified models simultaneously. The reduction procedure
we adopt has the unique property of being covariant
under rescaling of $\Sigma^{A B}$. By this we mean the following
 : let us imagine that somebody, using a particular
$\Sigma^{A B}$  and following our reduction scheme,
 arrives at a reduced equation, say (\ref{3.4}). Suppose
now that someone else multiplies (\ref{3.1}) with a function
$G(x)$. If he wants to reduce his scaled equation,
he has to find the $q$'s solving the system corresponding
to the conditions (\ref{3.2a}) with $\Sigma^{A B}$ replaced
by $G(x)\Sigma^{A B}$; it is therefore obvious that he will
 get the same set of $q$'s as the previous person. This
fact, together with the form of (\ref{3.3}), leads to the
conclusion that the reduced equation on $\mathcal{N}$ (for the
second person) will be just (\ref{3.4}) multiplied by
$g(q)=G(x)\Big|_{\mathcal{N}}$. This establishes the conformal
covariance of our reduction scheme. The form of
(\ref{3.2a}-\ref{3.2b}) makes also explicit the covariance of our
scheme under diffeomorphisms of $\mathcal{M}$ and $\mathcal{N}$.
\par \noindent
We would also like to briefly mention two important issues not
touched upon in this work:
\par \noindent
The first issue concerns the normalisability of the solutions to;
(\ref{2.7a}-\ref{2.7f}) and (\ref{2.10}); as it is well known,
$G^{\alpha \beta \gamma \delta}$ has signature $(-,+,+,+,+,+)$
which implies that the solutions to (\ref{2.7a}- \ref{2.7f}) are
not expected to be normalisable with respect to all
$\gamma_{\alpha \beta}$'s. $\Sigma^{A B}$ in (\ref{2.10}) can also
be seen to be of hyperbolic nature, implying that the same problem
will occur. This non-normalisability is intrinsically connected
with the long debated problem of time in Quantum Gravity and
Quantum Cosmology. In connection to this problem, our reduction
scheme (when it refers to one restricting condition) could be
taken as a basis for a satisfactory definition of probability on
any space-like
 hypersurface of $\Sigma^{A B}$; but clearly much work is
needed before something conclusive can be said.
Another
possibility of improving normalizability is the inclusion of
matter fields, mostly half integer spin fields \cite{zanelli}. In
particular the improvement is better if the matter is
supersymmetric \cite{death}. Recent results concerning
supersymmetric FRW \cite{obregon} and/or Bianchi models
\cite{tkach} show substantial improvement in the normalizability
of the wavefunctions.
\par \noindent
The second issue is related to the question of whether action
(\ref{2.4}) gives equations of motion equivalent to Einstein's
Field equations for the metric (\ref{2.2}); For unrestricted
$\gamma_{\alpha \beta}$ the answer is that the two sets
of equations are equivalent only for Class A spacetimes. The
examples cited at the end of section III do contain Class A
spacetimes and might therefore correspond to valid classical
actions. To see whether they actually do, one would have to
test them case by case for the different Bianchi types and the
restrictions adopted.
 We have not done this since it lied beyond the
scope of our work.
\par \noindent
Finally, we would like to remark that, for the case of Bianchi
type I, $x^{1}=x^{2}=0$ and equation (\ref{2.10}) reduces to an
Euler Equation for the single variable $x^{3}=\gamma$. This
indicates that the configuration space for this type is
effectively one-dimensional, a number that points to the
one-parameter Kasner family of classical solutions. The
explanation for this reduction of the seemingly 12-dimensional
phase-space (6 $\gamma_{\alpha \beta}$ plus 6 $\pi^{\alpha
\beta}$) involves considering the action of the 9-parameter
automorphism group (GL(3,R)) and is properly given by A. Ashtekar
and J. Samuel (see in \cite{jantzen}). The fact that no
$H_{\alpha}$'s exist (due to the vanishing of the structure
constants) is adequately compensated by the existence of many
classical integrals of motion; their quantum counterparts can be
used to reduce the degrees of freedom to $\gamma\equiv
det[\gamma_{\alpha\beta}]$ \cite{theo16}.
\Acknow {An early version of this work is part of the 1995 PENNED
program ``Quantum and Classical Gravity-Black holes'' (no 512)
supported by the General Secretariat for the Research and Industry
of the Hellenic Department of Industry, Research and Technology.}
\par
\noindent The authors wish to thank Professor A.B. Lahanas, Dr. G.
Diamandis, Dr. V. Georgalas and Dr. V.C. Spanos for helpful
discussions and Professor M.A.H. MacCallum for clarifying the
issue of automorphisms in connection to the reduction to $x^{A}$'s
(see appendix).
\par\noindent
One of the authors (E.C.V.) acknowledges financial
support from the University of Athens' Special Account for
Research.
\newpage
\section{Appendix : Calculating the Invariants}

We give an outline of how the relations between higher
order (in $C^{\alpha}_{\beta \gamma}$) scalars and
the basic variables ($x^{1},x^{2}, x^{3}$) arises.
\par \noindent
Since $C^{\alpha}_{\beta \gamma}$ is antisymmetric
in its lower indices, it can be expressed (in three
dimensions) in terms of the totally antisymmetric
 symbol $\epsilon_{\alpha \beta \gamma}$,
a symmetric matrix $m^{\alpha \beta}$ and a vector
$a_{\gamma}$ :
\begin{eqnarray*}
C^{\alpha}_{\beta \gamma}=\epsilon_{\beta \gamma \delta}
m^{\delta \alpha}+\delta^{\alpha}_{\gamma}a_{\beta}-
\delta^{\alpha}_{\beta}a_{\gamma}\Leftrightarrow
m^{\alpha \beta}=\frac{1}{2}C^{(\alpha}_{\gamma \delta}
\epsilon^{\beta)\gamma \delta}
\end{eqnarray*}
where the parenthesis means symmetrization :
\begin{eqnarray*}
a_{\gamma}=\frac{1}{2}C^{\alpha}_{\gamma \alpha}
\hspace{1cm}\mbox{and}\hspace{1cm}\epsilon
^{\alpha \beta \gamma}=\epsilon_{\alpha \beta \gamma}.
\end{eqnarray*}
Using this representation of $C^{\alpha}_{\beta \gamma}$
(or directly through an algebraic computing facility) one
can see that :
\begin{eqnarray*}
\gamma^{\nu \lambda}C^{\mu}_{\mu \nu}C^{\beta}_{\beta
\lambda}=\lambda \gamma^{\beta \nu}C^{\alpha}_
{\beta \delta}C^{delta}_{\nu \alpha}
\end{eqnarray*}
with the number $\lambda$ given in the following table :
\\
\begin{center}
\begin{tabular}{|c|c|c|c|c|c|c|c|c|} \hline
&&&&&&&&\\
Bianchi Type & II & III & IV & V & VI & VII & VIII & IX \\
&&&&&&&&\\
\hline
 & & & & & & & &  \\
$\lambda$ & any & 1 & 2 & 2 &
$\displaystyle{\frac{(h+1)^2}{h^{2}+1}}$ &
$\displaystyle{\frac{h^{2}}{h^{2}-2}}$ & 0 & 0 \\
&&&&&&&&\\
\hline
\end{tabular}
\end{center}
\par \noindent
\\
Under the (possibly time dependent) changes :
\begin{eqnarray*}
\sigma^{\alpha}\rightarrow\tilde{\sigma}^{\alpha}=
\Lambda^{\alpha}_{\beta}\sigma^{\beta}\end{eqnarray*}
where $\epsilon_{\alpha\beta \gamma}$
 is a tensor density of weight 1 and therefore
$m^{\alpha \beta}$ has weight -1 (so that $C^{\alpha}
_{\beta \gamma}$ be a tensor under GL(3,R)). This means
that $m:=det[m^{\alpha \beta}]$ is a scalar density
of weight -1 as well (m is non zero only for Bianchi
types VIII, IX). Since $\gamma_{\alpha \beta}$ is a second
rank tensor $\gamma := det[\gamma_{\alpha \beta}]$ is a
scalar density of weight -2.
\par \noindent
The calculation of the various scalars given below proceeds
as follows :
\par \noindent
One can use a matrix $\Lambda_{1}$\hspace{0.15cm}$\in$ GL(3,R) to
transform $\gamma_{\alpha \beta}$ to the identity matrix $I_{3}$.
Under such a transformation $m^{\alpha \beta}$ transforms to an
arbitrary symmetric matrix $\tilde{m}^ {\alpha \beta}$.
Subsequently one can use an orthogonal matrix $\Lambda_{2}$ to
diagonalise $\tilde{m}^{\alpha \beta}$ (keeping $\gamma_{\alpha
\beta}$ the identity matrix). The product
$\Lambda=\Lambda_{1}\Lambda_{2}$ transforms $\gamma_{\alpha
\beta}$
 to $I_{3}$ and puts $m^{\alpha \beta}$ to diagonal form.
\par \noindent
In this way we have the structure constants in terms of the three
eigenvalues of $m^{\alpha \beta}$ say $a, b, c$ and the
components $d, e, f$ of the vector $a_{\gamma}$. The Jacobi
identities imply that $a_{\gamma}$ is a null eigenvector of
 $m^{\alpha \beta}$ and thus $ad=be=cf=0$.
\par \noindent
In this representation one can (through an algebraic computing
facility) see that :
\begin{eqnarray*}
x^{1}&=&4\left(d^{2}+e^{2}+f^{2}\right)
+2\left(a^{2}+b^{2}+c^{2}\right)\\
x^{2}&=&2\left(d^{2}+e^{2}+f^{2}-ab-ac-bc\right)
\end{eqnarray*}
\par \noindent
and for example :
\begin{eqnarray*}
w_{1}:=C^{\iota}_{\xi \kappa}C^{\xi}_{\iota \lambda}
C^{\mu}_{\nu o}C^{\nu}_{\mu \pi}\gamma^{\kappa o}
\gamma^{\lambda \pi}=\left[2\left(d^2+e^2+f^2-ab-ac-bc
\right)\right]^2-8abc(a+b+c).
\end{eqnarray*}
\par \noindent
The first term in $w_{1}$ is $\left(x^2\right)^2$. The second
term has to be a scalar. Therefore although at first sight
it looks like -$8mm^{\alpha \beta}\gamma_{\alpha \beta}$ it
is actually the scalar- $\frac{8m}{x^{3}}m^{\alpha \beta}
\gamma_{\alpha \beta}$. Thus
\begin{eqnarray*}
w_{1}:=C^{\iota}_{\xi \kappa}C^{\xi}_{\iota \lambda}
C^{\mu}_{\nu o}C^{\nu}_{\mu \pi}\gamma^{\kappa o}
\gamma^{\lambda \pi}=\left(x^{2}\right)^{2}-
\frac{8m}{x^{3}}m^{\alpha \beta}\gamma_{\alpha \beta}
\end{eqnarray*}
\par \noindent
Working similarly, the following scalars (appearing when
one tries to calculate $\Sigma^{A B}$ and thus reach to
(\ref{2.10})) can be calculated.
\begin{eqnarray*}
w_{2}&:=&C^{\iota}_{\xi \kappa}C^{\xi}_{\iota \lambda}
C^{\mu}_{\nu o}C^{\nu}_{\mu \pi}\gamma^{\kappa o}
\gamma^{\lambda \pi}=\lambda\left(x^{2}\right)^{2}
\\
w_{3}&:=&C^{\xi}_{\xi \lambda}C^{\iota}_{\iota \kappa}
C^{\mu}_{\nu o}C^{\pi}_{\sigma \rho}\gamma^{\kappa \nu}
\gamma^{\lambda \sigma}\gamma^{o \rho}\gamma_{\mu \pi}
=\frac{\lambda x^{1}x^{2}}{2}
\\
w_{4}&:=&C^{\iota}_{\xi \kappa}C^{\xi}_{\iota \lambda}
C^{\kappa}_{\mu \nu}C^{\lambda}_{o \pi}\gamma^{\mu o}
\gamma^{\nu \pi}=-\frac{36m}{x^3}m^{\alpha \beta}\gamma
_{\alpha \beta}\\
w_{5}&:=&C^{\iota}_{\xi \kappa}C^{\xi}_{\iota \lambda}
C^{\rho}_{\mu \nu}C^{\sigma}_{o \pi}\gamma^{\kappa \mu}
\gamma^{\lambda o}\gamma^{\nu \pi}\gamma_{\rho \sigma}
=\frac{x^{1}x^{2}}{2}+\frac{2m}{x^{3}}m^{\alpha \beta}
\gamma_{\alpha \beta}\\
w_{6}&:=&C^{\iota}_{\xi \kappa}C^{\lambda}_{\mu \nu}
C^{o}_{\pi \theta}C^{\rho}_{\sigma \tau}\gamma^{\xi \mu}
\gamma^{\kappa \pi}\gamma^{\nu \sigma}\gamma^{\theta \tau}
\gamma_{\iota \lambda}\gamma_{o \rho}=\frac{(x^{1})^{2}}{2}-(\lambda-1)^{2}\frac{(x^{2})^{2}}
{2}+\frac{4m}{x^{3}}m^{\alpha \beta}
\gamma_{\alpha \beta}\\
w_{7}&:=&C^{\iota}_{\xi \kappa}C^{\lambda}_{\mu \nu}
C^{o}_{\pi \theta}C^{\rho}_{\sigma \tau}\gamma^{\xi \pi}
\gamma^{\mu \sigma}\gamma^{\iota \theta}\gamma^{\nu \tau}
\gamma_{\iota \lambda}\gamma_{o \rho}=(x^{1})^{2}-
2(\lambda-1)^{2}(x^{2})^{2}
+\frac{16m}{x^{3}}m^{\alpha \beta}
\gamma_{\alpha \beta}\\
w_{8}&:=&C^{\iota}_{\xi \kappa}C^{\lambda}_{\mu \nu}
C^{o}_{\iota \pi}C^{\sigma}_{\lambda \theta}\gamma^{\xi \mu}
\gamma^{\kappa \nu}\gamma^{\pi \theta}\gamma_{\sigma o}
=(1-\lambda^{2})(x^{2})^{2}
+\lambda x^{1}x^{2}+\frac{8m}{x^{3}}m^{\alpha \beta}
\gamma_{\alpha \beta}\\
w_{9}&:=&C^{\iota}_{\xi \kappa}C^{\xi}_{\iota \lambda}
C^{\mu}_{\nu o}C^{\pi}_{\theta \rho}\gamma^{\kappa \nu}
\gamma^{\lambda \theta}\gamma^{o \rho}\gamma_{\mu \pi}
=\frac{x^{1}x^{2}}{2}+\frac{2m}{x^{3}}m^{\alpha \beta}
\gamma_{\alpha \beta}
\end{eqnarray*}
\par \noindent
Using (\ref{2.8}) we find the curvature invariants to be :
\begin{eqnarray*}
R_{1}&=&x^{1}+2(1+2\lambda)x^{2}\\
R_{2}&=&3(x^{1})^{2}+4(4\lambda-1)(x^{2})^{2}+4x^{1}x^{2}+
64\frac{m}{x^{3}}m^{\alpha \beta}
\gamma_{\alpha \beta}\\
R_{3}&=&3(x^{1})^{3}+8(-4\lambda^{3}+12\lambda^{2}-
6\lambda+1)(x^{2})^{3}+6(1+2\lambda)(x^{1})^{2}x^{2}
+12(x^{2})^{2}x^{1}-\\
&&192\frac{m}{x^{3}}
\left(-\frac{m^{\alpha \beta}\gamma_{\alpha \beta}
m^{\gamma \delta}m_{\gamma \delta}}{x^{3}}+
\frac{2m^{\alpha \beta}m_{\beta \gamma}m^{\gamma}_{\alpha}}
{x^{3}}+2m\right)
\end{eqnarray*}
\par \noindent
It now remains to express the terms involving the
contractions of $m^{\alpha \beta}$ and $\gamma_
{\alpha \beta}$ as functions of $x^{1},x^{2},x^{3}$. It
is straightforward to see that :
\begin{eqnarray*}
\left(\frac{m^{\alpha \beta}\gamma_{\alpha \beta}}
{\sqrt{x^{3}}}\right)^{2}=(a+b+c)^{2}=\frac{x^{1}-2x^{2}}{2}
\Leftrightarrow \frac{m^{\alpha \beta}\gamma_{\alpha \beta}}
{\sqrt{x^{3}}}=\epsilon \sqrt{\frac{x^{1}-2x^{2}}{2}},
\hspace{0.5cm} \epsilon=sign(m^{\alpha \beta}
\gamma_{\alpha \beta})
\end{eqnarray*}
\par \noindent
Analogously one finds that :
\begin{eqnarray*}
\frac{m^{\gamma \delta}m_{\gamma \delta}}{x^{4}}=
\frac{x^{1}-\lambda x^{2}}{2}\hspace{0.5cm}
\mbox{and}\hspace{0.5cm}\frac{m^{\alpha \beta}
m_{\beta \gamma}m^{\gamma}_{\alpha}}{x^{3}\sqrt{x^{3}}}=
\epsilon\sqrt{\frac{x^{1}-2x^{2}}{2}}\left(
\frac{2x^{1}+(2-3\lambda)x^{2}}{4}\right)+\frac{3m}
{\sqrt{x^{3}}}
\end{eqnarray*}


\begin{thebibliography}{99}
\bibitem{theo3} This is only meant as a working hypothesis
for demonstration of the argument; there are claims in the literature
that the smooth functionals are a set of measure zero in the space of all
functionals. However, in the absence of a mathematically rigorous
 measure theory for the space of functionals, we feel that one is allowed
to make such a hypothesis.
\end{thebibliography}
 \end{document}